\documentclass[superscriptaddress,showkeys,notitlepage,onecolumn]{revtex4-1}%
\usepackage{amsmath,amssymb,amsbsy,subfigure,hyperref,subfigure,bbm,times,txfonts,accents,float}
\usepackage{caption}
\usepackage{graphicx}
\usepackage{dcolumn}
\usepackage{bm}
\usepackage{lineno,hyperref}
\usepackage{graphicx}
\usepackage{epstopdf}
\modulolinenumbers[5]
\emergencystretch=\maxdimen
\hypersetup{colorlinks=true,linkcolor=blue,anchorcolor=blue,citecolor=blue}

\begin{document}

\title{Dynamic Investigation of the New Quantum-Control-Assisted Reverse uncertainty relation}

\author{Qiyi Li}
\affiliation{%
School of Physics, Beihang University, South Third Street No.9, Beijing 102206, China
}%

\author{Shaoqiang Ma}
\affiliation{%
	Jinhang Digital Technology Co., Ltd, Jingshun Road No.7, Beijing 100028, China
}%

\author{Sansheng Wang}
\affiliation{%
	School of Physics, Beihang University, South Third Street No.9, Beijing 102206, China
}%
\author{Xiao Zheng}
\email{xiaozheng@buaa.edu.cn}
\affiliation{%
	School of Physics, Beihang University, South Third Street No.9, Beijing 102206, China
}%
\author{Guofeng Zhang}
\email{gf1978zhang@buaa.edu.cn}
\affiliation{%
	School of Physics, Beihang University, South Third Street No.9, Beijing 102206, China
}%

\date{\today}

\begin{abstract}
Recently, a new interesting concept of reverse uncertainty relation is introduced. Different from the normal uncertainty relation, the reverse one indicates that one cannot only prepare quantum states with joint small uncertainty, but also with joint great uncertainty for incompatible observables. We in this work construct a new quantum-control-assisted reverse uncertainty relation and investigate the corresponding dynamic evolution in the Heisenberg model with Dzyaloshinskii-Moriya interaction. The obtained relation indicates that the reverse uncertainty can be broken with help of the quantum control system. The dynamic investigation reveals that there exists an interesting single-value relationship between new uncertainty relation and the mixedness of the system, indicating that the tightness and upper bound of the uncertainty relation can be written as functional form of the mixedness. By comparing the existing research in [Physica Scripta 2023, 98(6), 065113], we show that the single-value relationship with the mixedness is the common nature of both the normal uncertainty relations and the reverse uncertainty relation.\\

\end{abstract}
\maketitle
\section{Introduction}
\textbf{Quantum uncertainty relation, indicating that one cannot simultaneously and precisely predict two incompatible observables\cite{1,2,3,4,5,003,005,004}, is an inherent characteristic of quantum systems and plays crucial roles in the field of quantum information science such as, quantum communication\cite{9,20,21,22,23,24,26,27,28,29}, quantum computing\cite{010,25} and quantum phase transition\cite{008,25}. A number of uncertainty relations have been construted, which has greatly promoted the development of the above fields.} The most fundamental and famous uncertainty relation is the Schrödinger uncertainty relation (SUR), which reads\cite{3}:
\begin{align}
    \tag{1}
    \bigtriangleup \hat{A}^{2} \bigtriangleup \hat{B}^{2} \ge \frac{1}{4} \left | \left \langle \left [ \hat{A},\hat{B} \right ] \right \rangle  \right | ^{2}+\frac{1}{4} \left | \left \langle \left \{ \check{A} ,\check{B} \right \} \right \rangle  \right | ^{2} \nonumber
\end{align}
\textbf{where $\bigtriangleup \hat{A}^{2}(\bigtriangleup \hat{B}^{2})$ represents the variance of the observable $\hat{A}(\hat{B})$, $\check{A}=\hat{A}-\left \langle \hat{A} \right \rangle (\check{B}=\hat{B}-\left \langle \hat{B} \right \rangle)$, $\left [ \hat{A},\hat{B} \right ]=\hat{A}\hat{B}-\hat{B}\hat{A}$ and $\left \{ \check{A},\check{B}  \right \}  = \check{A}\check{B}+\check{B}\check{A}$. $\left \langle \hat{A} \right \rangle (\left \langle \hat{B} \right \rangle)$ is the expectation of $\hat{A} (\hat{B})$, i.e., $\left \langle \hat{A} \right \rangle (\left \langle \hat{B} \right \rangle) = Tr(\rho \hat{A}(\hat{B}))$ with $\rho$ being the density matrix of the system.} However, the uncertainty relation in product form, for instance SUR, will suffer the triviality problem, i.e., the corresponding lower bound equals to zero even for incompatible observables\cite{36,37,44,45}. To fix the trivial problem, Maccone and Pati constructed a new variance-based uncertainty relation in sum form\cite{44}:
\begin{align}
    \tag{2}
    \bigtriangleup \hat{A}^{2} +\bigtriangleup \hat{B}^{2} = \pm i \left \langle \left [ \hat{A},\hat{B} \right ]  \right \rangle +\left | \left \langle \psi \left | \hat{A} \pm i\hat{B} \right |  \psi^{\bot} \right \rangle  \right | ^{2} \nonumber
\end{align}
where $\left | \psi \right \rangle$ represents the pure state of the system, and $\left | \psi^{\bot} \right \rangle$ is the corresponding orthogonal state. However, due to the introduction of the orthogonal state, it is difficult to apply the uncertainty relation (2) to high-dimensional situation\cite{36}. To solve above problems, Ref.\cite{36a} constructed a unified and exact framework for the variance-based uncertainty relations by introducing a new concept of the auxiliary operator:\\
\begin{align}
    \tag{3}
    \left \langle \mathcal{A}^{\dagger}\mathcal{A} \right \rangle \left \langle \mathcal{B}^{\dagger}\mathcal{B} \right \rangle=\frac{1}{4}\left | i\left \langle \left [ \mathcal{A},\mathcal{B} \right ]_{GC} \right \rangle \right |^{2}+\frac{1}{4}\left | i\left \langle \left \{ \mathcal{A},\mathcal{B} \right \}_{GC} \right \rangle \right |^{2}+\left \langle \mathcal{B}^{\dagger}\mathcal{B} \right \rangle\left \langle \mathcal{C}^{\dagger}\mathcal{C} \right \rangle \nonumber
\end{align}
where $\mathcal{A}$ and $\mathcal{B}$ are two incompatible observables and $\mathcal{C} = \mathcal{A}- \left \langle \mathcal{B}^{\dagger} \mathcal{A} \right \rangle \mathcal{B}  /\left \langle \mathcal{B}^{\dagger}\mathcal{B} \right \rangle$. $\left \langle \mathcal{Q}^{\dagger}\mathcal{Q} \right \rangle$ represents the second order origin moment of the observable $\mathcal{Q}$, $\left [ \mathcal{A},\mathcal{B} \right ]_{GC} = \mathcal{A}^{\dagger}\mathcal{B} - \mathcal{B}^{\dagger}\mathcal{A}$  and  $\left \{ \mathcal{A},\mathcal{B} \right \}_{GC} = \mathcal{A}^{\dagger}\mathcal{B} + \mathcal{B}^{\dagger}\mathcal{A}$. This unified framework can be used to solve the mentioned triviality problems by constructing a stronger uncertainty relation in sum form for two and more incompatible observables with the stronger lower bound\cite{36a}.\\

As we know, the uncertainty for the incompatible observables can also be measured by entropy. Therefore, the entropic uncertainty relation has been constructed and received extensive attention\cite{000,002,003}. One famous entropic uncertainty relation reads:
\begin{align}
    \tag{4}
    H(R)+H(S) \ge \log_{2}{(\frac{1}{c})}
    \nonumber
\end{align}
\textbf{where $R$ and $S$ are two incompatible observables. $H(R)(H(S))$ is the Shannon entropy of the observable $R(S)$, i.e., $H(R)(H(S))=-\sum p(\left | \varphi_{r} \right \rangle (\left | \phi_{s} \right \rangle))\log_{2}{p(\left | \varphi_{r} \right \rangle (\left | \phi_{s} \right \rangle))}$ with $\left | \varphi _{r}  \right \rangle (\left | \phi_{s} \right \rangle )$ represents the normalized eigenvector of $R(S)$, and $p(\left | \varphi_{r} \right \rangle (\left | \phi_{s} \right \rangle))$ is the probability that the measurement result is $\left | \varphi_{r} \right \rangle (\left | \phi_{s} \right \rangle)$ when $R(S)$ is performed.} $c = max_{r,s}\left | \left \langle \varphi_{r} \phi_{s} \right \rangle  \right | ^{2}$\cite{47}.\\

In order to break the limitation imposed by the aforementioned uncertainty relations, the uncertainty relation with additional conditions will be introduced subsequently. For the clarity of subsequent differentiation, uncertainty relations (1)-(4) are denoted here as the uncertainty relation without conditional system. In 2010, Berta $et$ $al$ constructed the quantum memory-assisted entropic uncertainty relation (QM-EUR), showing that the limitations of uncertainty relation without conditional system can be broken with the help of the quantum memory system\cite{6}. The QM-EUR reads:\\ 
\begin{align}
    \tag{5}
    H(R|B)+H(S|B) \ge \log_{2}{\frac{1}{c}}+H(A|B) \nonumber
\end{align}
\textbf{where $A$ stands for the measured system and $B$ represents the memory system. $H(A|B)=H(\rho_{AB})-H(\rho_{B})$ is the conditional von Neumann entropy of the density operator $\rho_{AB}$ . $\rho_{AB}$ represents the state of the whole system and $H(\rho)$ is the Shannon entropy of the state $\rho$. $\rho_{B}=Tr_{A}(\rho_{AB})$ represents the density operator of the memory system. $H(A|B)$ is actually a negative conditional entropy which represents a signature of entanglement between $A$ and $B$. It is therefore breaks the limit of uncertainty relation without conditional system\cite{6}. $H(Y|B)=H(\rho_{YB})-H(\rho_{B})$ represents the conditional von Neumann entropy of the density operator $\rho_{YB}$ with $\rho_{YB} = {\textstyle \sum_{Y}^{}} (\left | \varphi_{Y} \right > \left < \varphi_{Y} \right | \otimes I)\rho_{AB}(\left | \varphi_{Y} \right > \left < \varphi_{Y} \right | \otimes I)$ being the post-measurement state. $\left | \varphi_{Y} \right >$ is the eigenvector of $Y$ with $Y\in (R,S)$, and $I$ stands for identical operator.} $H(R|B)(H(S|B))$ stands for the uncertainty of the measurement $R(S)$ conditioned on the prior information stored in memory system $B$.\\

The QM-EUR is essentially a bipartite conditional normal uncertainty relation and it is a considerable challenge to extend this relation to multipartite situation\cite{00}. To fix this problem, Ref.\cite{00} constructed a new variance-based quantum-control-assisted multipartite uncertainty relation by introducing the conditional variance:\\
\begin{align} 
    \tag{6}
    \sum_{k=1}^{K}E\left [ V(Q_{k}^{A}|O_{k}^{C_{1}},\cdot \cdot \cdot ,O_{k}^{C_{n}}) \right ]\ge L_{tra}-\sum_{k=1}^{K}V\left [ E(Q_{k}^{A}|Q_{k}^{C_{1}}) \right ] \nonumber \\
    -\sum_{k=1}^{K} \sum_{n=2}^{N}E\left [ V(E\left [ Q_{k}^{A}|O_{k}^{C_{n}} \right ]|O_{k}^{C_{1}},\cdot \cdot \cdot O_{k}^{C_{n-1}}) \right ] \nonumber
\end{align}
where $O_{k}^{C_{n}}(Q_{k}^{A})$ represents the measurement $O_{k}(Q_{k})$ performed to particle $C_{n}(A)$. $E\left [ V(Q_{k}^{A}|O_{k}^{C_{1}},\cdot \cdot \cdot ,O_{k}^{C_{n}}) \right ]$ is the conditional variance of $Q_{k}^{A}$ when one has performed the measurements $O_{k}$ to particles $C_{1},\cdot \cdot \cdot ,C_{n}$. $V\left [ E(Q_{k}^{A}|Q_{k}^{C_{1}}) \right ] $ is the variance of $E({Q_{k}^{A}})$ under the condition that the measurement $O_{k}^{C_{1}}$ has been performed. $E\left [ V(E\left [ Q_{k}^{A}|O_{k}^{C_{n}} \right ]|O_{k}^{C_{1}},\cdot \cdot \cdot O_{k}^{C_{n-1}}) \right ] $ is defined as the conditional variance of $E(Q_{k}^{A}|O_{k}^{C_{n}})$ when one has performed the measurements $O_{k}^{C_{1}},\cdot \cdot \cdot O_{k}^{C_{n-1}}$. The quantum-control-assisted uncertainty relation has been extensively developed as is was successfully extended to multipartite case\cite{00,001}, thereby fostering the advancement of uncertainty relations.\\

The quantum memory system and quantum control system in uncertainty relations (5) and (6) are essentially conditional systems as they introduce memory system and control system entangled with measured system to add conditions. Uncertainty relations (5) and (6) are thereby distinguish from uncertainty relations (1)-(4). For the convenience of subsequent differentiation, uncertainty relations (5) and (6) are named here the conditional uncertainty relation\cite{003}.\\

Additionally, Debasis Mondal demonstrated there is an emerging concept known as the reverse uncertainty relation, which provides an upper bound for the uncertainty of the measurements for the incompatible observables\cite{36}. Similar to the normal uncertainty relation, the reverse one is also a unique feature of quantum mechanics. The normal uncertainty relation expresses the lower bound of the variance for the incompatible observables and the reverse uncertainty relation suggests the impossibility to make the variance of incompatible observables arbitrarily great concurrently. The first reverse uncertainty relation constructed by Debasis Mondal reads\cite{36}:
\begin{align}
    \tag{7}
    \bigtriangleup \hat{A}^{2}+\bigtriangleup \hat{B}^{2} \le \frac{2\bigtriangleup(\hat{A}-\hat{B})^{2}}{1-\frac{cov(\hat{A},\hat{B})}{\bigtriangleup \hat{A}.\bigtriangleup \hat{B}}}-2\bigtriangleup \hat{A}.\bigtriangleup \hat{B} \nonumber
\end{align}
where $cov(\hat{A},\hat{B})=\left \langle \left \{ \hat{A},\hat{B} \right \} \right \rangle /2-\left \langle \hat{A} \right \rangle \left \langle \hat{B} \right \rangle$. However, the reverse uncertainty relation (7) is trivial in some situations, i.e., the upper bound of uncertainty relation (7) approaches infinity when $\bigtriangleup \hat{A}.\bigtriangleup \hat{B}$ is precisely equal to $cov(\hat{A},\hat{B})$, which is meaningless in the physical picture. To solve this problem, Ref.\cite{02} constructed a series of new reverse uncertainty relation:\\
\begin{align}
    \tag{8}
    \bigtriangleup \hat{A}^{2}+\bigtriangleup \hat{B}^{2} \le \sqrt{Tr(\rho_{s}^{2})} \sqrt{Tr((\check{A}\pm i\check{B} )(\check{A}\mp  i\check{B} )(\check{A}\pm i\check{B} )(\check{A}\mp  i\check{B} ))}\pm i \left \langle[ \hat{A},\hat{B} \right ]\rangle \nonumber 
\end{align}

\begin{align}
    \tag{9}
    \bigtriangleup \hat{A}^{2}+\bigtriangleup \hat{B}^{2} \le \sqrt{Tr(\rho_{s}^{2})} Tr((\check{A}\pm i\check{B} )(\check{A}\mp  i\check{B} ))\pm i \left \langle[ \hat{A},\hat{B} \right ]\rangle \nonumber
\end{align}

\begin{align}
    \tag{10}
    \sum_{j=1}^{N}\bigtriangleup \hat{A}_{j}^{2} \le -\sum_{k<l} \left \langle \left \{ e^{i\theta}\check{A}_{k},e^{i\theta}\check{A}_{l} \right \}  \right \rangle +\sqrt{Tr(\rho _{s}^{2})(Tr\left [ (\sum_{k=1}^{N}e^{-i\theta}\check{A}_{k} )(\sum_{l=1}^{N}e^{i\theta}\check{A}_{l})(\sum_{k=1}^{N}e^{-i\theta}\check{A}_{k} )(\sum_{l=1}^{N}e^{i\theta}\check{A}_{l}) \right ] -M )}  \nonumber
\end{align}
where $\theta$ is a value decided by the system density matrix $\rho$ and $\theta \in \left [ 0,2\pi \right ]$. $M=(| \left \langle [F_{2},O_{2}] \right \rangle |^{2}+| \left \{ [F_{2},O_{2}] \right \} |^{2})/4|\left \langle O_{2}^{T}O_{2} \right \rangle|$ represents the embodiment of the auxiliary operator in the reverse uncertainty relation\cite{02}, with $F_{2}=F_{1}-\left \langle O_{1}F_{1} \right \rangle \cdot O_{1}/|\left \langle O_{1}^{T}O_{1} \right \rangle |$ , $F_{1}=(e^{i\theta}\check{A}+e^{i\theta}\check{B})+(e^{i\theta}\check{B}+e^{i\theta}\check{A})$ and $O_{k}$ being the auxiliary operator. $O_{1}$ is here taken as the density matrix. \\

One can observe that the uncertainty relations (7), (8), (9) and (10) are distinguish from aforementioned uncertainty relations as they expresses the upper bound instead of lower bound for the uncertainty. They are therefore called here the reverse uncertainty relation without conditional system for convenient differentiation subsequently.\\

As mentioned above, the uncertainty relation without conditional system can be broken by the conditional one, and a natural question arises that can the reverse uncertainty relation without conditional system be broken by the conditional reverse one, so as to achieve the purpose of accurately measuring incompatible observables in the same time.\\

In this work, we construct a new quantum-control-assisted reverse uncertainty relation and investigate the dynamic evolution of this relation in the Heisenberg system with the Dzyaloshinskii-Moriya interaction. The obtained relation suggests that the reverse uncertainty can be broken with the help of a quantum control system. The dynamic investigation reveals an interesting single-valued relationship between the new uncertainty relation and the mixedness of the system, indicating the tightness and upper bound of the uncertainty relation exhibits high degree of correlation with the mixedness of the system. This single-value relationship is found as the common nature of both the normal uncertainty relation and the reverse uncertainty relation through further comparison. The following article is divided into four sections. Sec.II is the construction of the new quantum-control-assisted reverse uncertainty relation. In Sec.III, the dynamic evolution in the Heisenberg model with DM interaction is investigated. Finally, Sec.IV is devoted to the conclusion.\\


\section{construction of conditional reverse Uncertainty Relation} \label{C2}
The new quantum-control-assisted reverse uncertainty relation for K incompatible observables reads:\\
\begin{align}
    \tag{11}
	\sum_{k=1}^K\mathrm {E}\left[\mathrm {V}\left(Q_k^A|O_k^{C}\right)\right]\leq U_{tra}-\sum_{k=1}^K\mathrm {V}\left[\mathrm {E}\left(Q_k^A|O_k^{C}\right)\right] \nonumber.
\end{align}
where $O_k$ stands for an arbitrary observable with $k=1,2,\cdots,K$, and $\mathrm {E}\left[\mathrm {V}\left(Q_k^A|O_k^{C}\right)\right]$ is the conditional variance of $Q_k^A$ on the condition that we have performed the measurements $O_k^{C}$ . $\mathrm {V}\left[\mathrm {E}\left(Q_k^A|O_k^{C}\right)\right]$ represents the variance of $\mathrm {E}\left(Q_k^A|O_k^{C}:=\lambda^{(k)}_{j}\right)$ in terms of the probability $\mathrm {P}\left(O_k^{C}:=\lambda^{(k)}_{j}\right)$ with $\lambda^{(k)}_{j}$ being an eigenvalue of $O_k^{C}$, and $\mathrm {P}\left(O_k^{C}:=\lambda^{(k)}_{j}\right)$ is the probability that the measurement result is $\lambda^{(k)}_{j}$ when we perform the measurement $O_k^{C}$ on the subsystem $C$. $U_{tra}$ is the traditional upper bound which is taken as the upper bound of Eq.(8), (9) or (10).\\

To prove the reverse uncertainty relation (11), we have to prove the following equality first:\\
\begin{align}
    \tag{12}
    \mathrm{E}\left(Q^{A}\right)=\mathrm{E}\left[\mathrm{E}\left(Q^{A}|O^{{C}}\right)\right]\nonumber,
\end{align}
where $\mathrm{E}\left[\mathrm{E}\left(Q^{A}|O^{C}\right)\right]=\sum_{\lambda_j}P\left(\rho,O^{C}:=\lambda_j\right)\mathrm{E}\left(Q^{A}|O^{C}:=\lambda_j\right)$ represents the expectation of $\mathrm{E}\left(Q^{A}|O^{C}:=\lambda_j\right)$ in terms of the probability $P\left(\rho,O^{C}:=\lambda_j\right)$, with $\mathrm{E}\left(Q^{A}|O^{C}:=\lambda_j\right)=\mathrm{Tr}\left[Q^{A}\mathrm{Tr}^{C}\left(\rho_{(O^{C}:=\lambda_j )}\right)\right]$ being the conditional expectation of $Q^{A}$ given that the measurement result of $O^{C}$ is $\lambda_j$. $\operatorname{Tr}^{C}\left(\rho_{\left(O^{C}:= \lambda_j\right)}\right)$ represents the partial trace of $\rho_{\left(O^{C}:= \lambda_j\right)}$ over the basis of the system $C$.\\

Consider a system containing two subsystems $A$ and $C$, and assume the state of the whole system is $\rho$. We can obtain that:\\
\begin{align}
    \tag{13}
    P\left(\rho,O^{C}:=\lambda_j\right)=\operatorname {Tr}\left[ \left(I^{A}\otimes|\lambda_j\rangle^{C} \langle\lambda_{j}|^{C}\right)\rho \left(I^{A}\otimes |\lambda_j\rangle^{C}\langle\lambda_j|^{C}\right)\right]
\end{align}
where $I^A(I^C)$ is the identity operator of the system $A(C)$. The state of the whole system after the measurement turns into:\\
\begin{align}
    \tag{14}
    \rho_{\left(O^{C}:=\lambda_j\right)}=\frac{\left(I^{A}\otimes|\lambda_j\rangle^{C} \langle\lambda_{j}|^{C}\right)\rho \left(I^{A}\otimes |\lambda_j\rangle^{C}\langle\lambda_j|^{C}\right)} {\operatorname {Tr}\left[ \left(I^{A}\otimes|\lambda_j\rangle^{C} \langle\lambda_{j}|^{C}\right)\rho \left(I^{A}\otimes |\lambda_j\rangle^{C}\langle\lambda_j|^{C}\right)\right]} \nonumber.
\end{align}

\begin{align}
    \tag{15}
    \operatorname{Tr}^{C}\left(\rho_{\left(O^{C}:= \lambda_j\right)}\right)=\frac{ \langle\lambda_{j}|^{C}\rho  |\lambda_j\rangle^{C}} {\operatorname {Tr}\left[ \left(I^{A}\otimes|\lambda_j\rangle^{C} \langle\lambda_{j}|^{C}\right)\rho \left(I^{A}\otimes |\lambda_j\rangle^{C}\langle\lambda_j|^{C}\right)\right]} \nonumber.
\end{align}

Then, one can obtain:\\
\begin{align}
    \tag{16}
    \mathrm{E}\left[\mathrm{E}\left(Q^{A}|O^{C}\right)\right]
    &=\sum_{\lambda_{j}}P\left(\rho,O^{C}:=\lambda_j\right)\mathrm{E}\left(Q^{A}|O^{C}:=\lambda_j \right)\nonumber \\
    &=\sum_{\lambda_{j}}\operatorname {Tr}\left[ \left(I^{A}\otimes|\lambda_j\rangle^{C} \langle\lambda_{j}|^{C}\right)\rho \left(I^{A}\otimes |\lambda_j\rangle^{C}\langle\lambda_j|^{C}\right)\right]\operatorname{Tr}\left[Q^{A}\operatorname{Tr}^{C}\left(\rho_{\left(O^{C}:= \lambda_j\right)}\right)\right]\nonumber \\
    &=\sum_{\mu_i}\langle\mu_i|^{A}Q^{A}\left(\sum_{\lambda_j}\langle\lambda_j|^{C}\rho|\lambda_j\rangle^{C}\right)|\mu_i\rangle^{A}\nonumber \\
    &= \operatorname{Tr}\left[Q^{A}\operatorname{Tr}^{C}\left(\rho\right)\right]=\mathrm{E}\left(Q^{A}\right)\nonumber.
\end{align}

Thus the proof of Eq.(12) is completed. One can obtain that:\\
\begin{align}
    \tag{17}
    \mathrm{V}\left(Q^A\right)&=\mathrm{E}\left({Q^{A}}^2\right)-\left[\mathrm{E}\left(Q^{A}\right)\right]^2 \nonumber \\
    &=\mathrm{E}\left[\mathrm{E}\left({Q^A}^2|O^{C}\right)\right]-\left\{\mathrm{E}\left[\mathrm{E}\left(Q^{A}|O^{C}\right)\right] \right\}^{2}\nonumber \\
    &=\mathrm{E}\left[\mathrm{V}\left(Q^A|O^{C}\right)\right]+\mathrm{E}\left[\mathrm{E}\left(Q^{A}|O^{C}\right)^{2}\right]-\left\{\mathrm{E}\left[\mathrm {E}\left(Q^{A}|O^{C}\right)\right]\right\}^{2}\nonumber \\
    &=\mathrm{E}\left[\mathrm{V}\left(Q^{A}|O^{C}\right)\right]+\mathrm{V}\left[\mathrm{E}\left(Q^{A}|O^{C}\right)\right]\nonumber
\end{align}

Then, combining Eq.(8), (9) and (10), one can obtain the conditional reverse uncertainty relation based on the control system. The constructed conditional reverse uncertainty relation can be described by following games: (i). Bob prepares two entangled particles, $A$ and $C$. $A$ is the measured system and $C$ represents the control system; (ii). Bob sends the measured system $A$ to Alice; (iii). Alice selects a measurement to be performed, such as $Q_{k}$, and tells it to Bob; (iv). With the information Bob has about the quantum state of the whole system, an appropriate measurement, such as $O_{k}$, has been performed by Bob to the control system $C$; (v). Alice performs $Q_{k}$ on the measured system $A$.\\

\begin{figure}[H]
    \centering
    \resizebox{14cm}{5cm}{\includegraphics{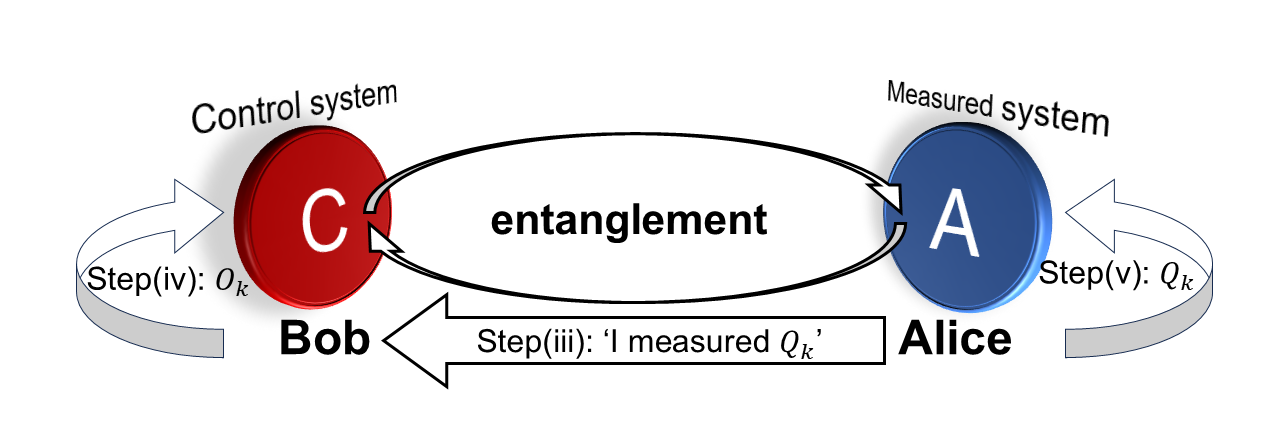}}
    \caption*{Fig.1 Schematic diagram of quantum-control-assisted reverse uncertainty relation}\label{Fig0}
\end{figure}

In the above games, the measurements performed on the control systems $C$ are called quantum control. Actually, the second term in the right-hand side of uncertainty relation (11) quantifies the quantum control. This term is greater than 0 when $A$ and $C$ are entangled, thus achieving the purpose of breaking the limit of the reverse uncertainty relation without conditional system. This is to say, the quantum control can be used to reduce the uncertainty of the measured system $A$ when $A$ and $C$ are two entangled particles.\\

\section{The Dynamic Evolution of the reverse uncertainty relation} \label{C3}
In this section, we will focus on the dynamic evolution analysis. \textbf{To start with, the Hamiltonian of the Heisenberg model with DM interaction of two particles is\cite{49,50}:} \\
\begin{align}
    \tag{18}
     H_{DM}= \frac{J}{2}[\sigma_{1x}\sigma_{2x}+\sigma_{1y}\sigma_{2y}+\sigma_{1z}\sigma_{2z}+D(\sigma_{1x}\sigma_{2y}-\sigma_{1y}\sigma_{2x})] \nonumber
    \\
    =J[(1+iD)\sigma_{1+}\sigma_{2-}+(1-iD)\sigma_{1-}\sigma_{2+}] \nonumber
\end{align}
where $J$ is the real coupling coefficient. $\vec{D}=\vec{D_{z}} $ is the DM coupling vector\cite{49,50}. \textbf{$\sigma_{ij}$ denotes the $i$-direction pauli operator acting on particle $j$ with $i = x, y, z$ and $j = 1, 2$. $\sigma_{1+}$ and $\sigma_{1-}$ represent the raising operator and lowering operator performed on particle 1, respectively. Similarly, $\sigma_{2+}$ and $\sigma_{2-}$ represent the raising and lowering operators for particle 2.} The quantum state of the system becomes $\rho(T)=e^{-\beta H}/Z$ as the system reaches the thermal equilibrium state\cite{007}. $H$ is the Hamiltonian of the system, $T$ represents the temperature of the system, and $Z=Tr(e^{-\beta H})$ is the partition function. \textbf{$\beta =1/k_{B}T$ with Boltzmann constant being taken as $k_{B} = 1$ for simplification.} Combining what mentioned above, the density matrix $\rho(T)$ of the Heisenberg model with DM interaction in thermal equilibrium state can be expressed as\cite{48}:\\
\begin{align}
    \tag{19}
    \rho(T)=\frac{1}{Z}
    \begin{pmatrix}
        \rho_{11} & 0 & 0 & 0 \\
        0 & \rho_{22} & \rho_{23} & 0 \\
        0 & \rho_{32} & \rho_{33} & 0 \\
        0 & 0 & 0 & \rho_{44} \\
    \end{pmatrix}
\end{align}
with the elements $\rho_{11}=\rho_{44}=e^{-\beta J/2}$, $\rho_{22}=\rho_{33}=e^{\beta (J-\delta)/2}(1+e^{-\beta \delta})/2$, $\rho_{23}=e^{i\theta}e^{\beta (J-\delta)/2}(1-e^{\beta \delta})/2$, $\rho_{32}=e^{-i\theta}e^{\beta (J-\delta)/2}(1-e^{\beta \delta})/2$, $Z=2e^{-\beta J/2}(1+e^{\beta J}\cosh (\beta \delta /2))$,  and $\delta =2J\sqrt{1+D^{2}}$.\\

In the following, the concurrence $C$ is used to measure the entanglement of the two qubits Heisenberg system with DM interaction. $C$ reads\cite{51}:\\
\begin{align}
    \tag{20}
    C=\frac{2}{Z}\mathrm{max}\left [ \frac{1}{2}|e^{\frac{\beta (J-\delta)}{2}}(1-e^{\beta \delta})|-e^{\frac{\beta J}{2}},0 \right ]\nonumber
\end{align}

According to Ref.\cite{01}, \cite{03} and \cite{04}, $1-Tr(\rho^{2})$ is positively correlated with the mixedness of $\rho$. Therefore, we introduced $\gamma =1-Tr(\rho^{2} )$ to measure the mixedness of the system:\\
\begin{align}
    \tag{21}
    \gamma =\frac{4e^{\beta (J+\delta )}[\cosh (\beta J)+2\cosh (\frac{\beta \delta}{2})]}{[e^{\beta (J+\delta )}+e^{\beta J}+2e^{\frac{\beta \delta}{2}}]^{2}}\nonumber
\end{align}

Then, $W$ and $U$ are introduced to measure the performance and the tightness of the conditional reverse uncertainty relation, which read\cite{006}:\\
\begin{align}
    \tag{22}
    W=U_{tra}-\sum_{k=1}^{K}V[E(Q_{k}^{A}|O_{k}^{C})]\nonumber
\end{align}


\begin{align}
    \tag{23}
    U=\frac{W}{\sum_{k=1}^{K}E[V(Q_{k}^{A}|O_{k}^{C})]}\nonumber
\end{align}
where $W$ is the upper bound of uncertainty relation (11) and $U$ is the ratio of the left to right side of uncertainty relation (11). Obviously, the smaller the value of $W$ is, the better the performance of the uncertainty relation is. And the closer the value of $U$ to 1, the better the tightness of the system is. In the following, we will respectively investigate the dynamic evolution of the conditional reverse uncertainty relation from the perspective of the upper bound $W$ and the tightness $U$.\\

\begin{figure}[H]
    \centering
    \subfigure[Evolution of $C$ with $J$ and $D$ when $T = 0.5$]{%
        \resizebox{6cm}{4cm}{\includegraphics{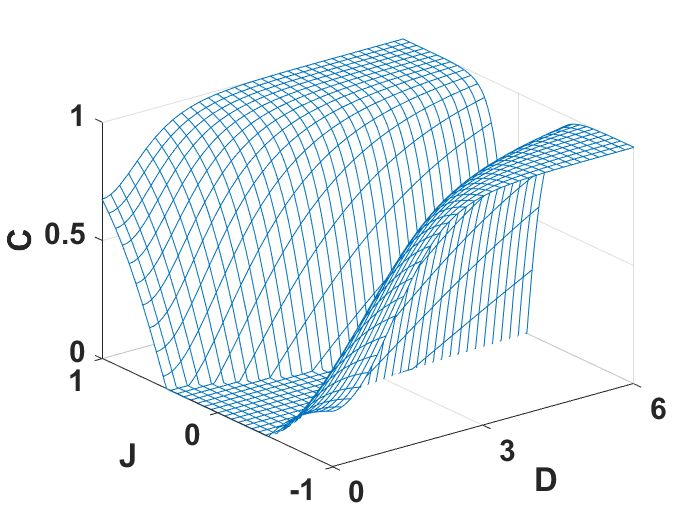}}}
    \hspace{5pt}
    \subfigure[Evolution of $C$ with $J$ and $D$ when $T = 1$]{%
        \resizebox{6cm}{4cm}{\includegraphics{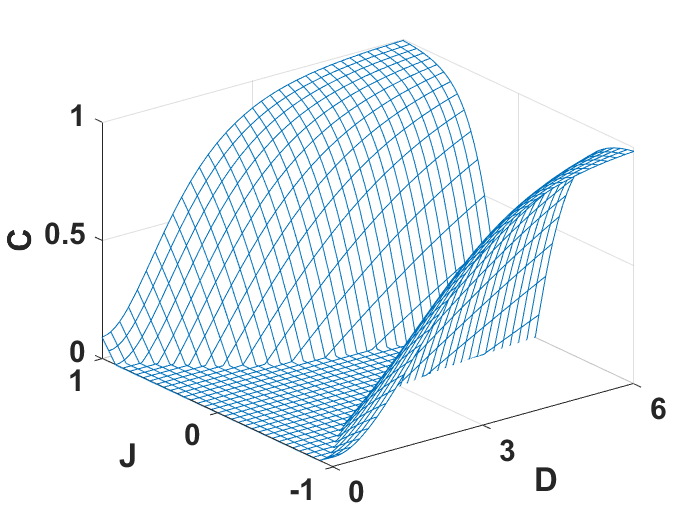}}}
    \vfill
    \subfigure[Evolution of $\gamma$ with $J$ and $D$ when $T = 0.5$]{%
        \resizebox{6cm}{4cm}{\includegraphics{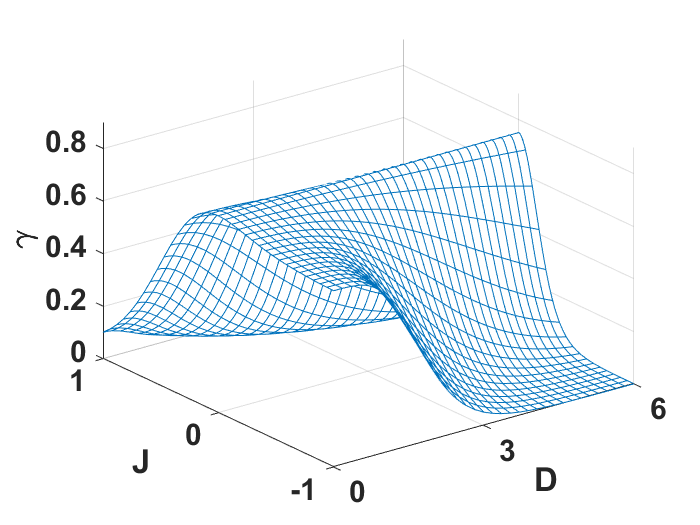}}}
    \hspace{5pt}
    \subfigure[Evolution of $\gamma$ with $J$ and $D$ when $T = 1$]{%
        \resizebox{6cm}{4cm}{\includegraphics{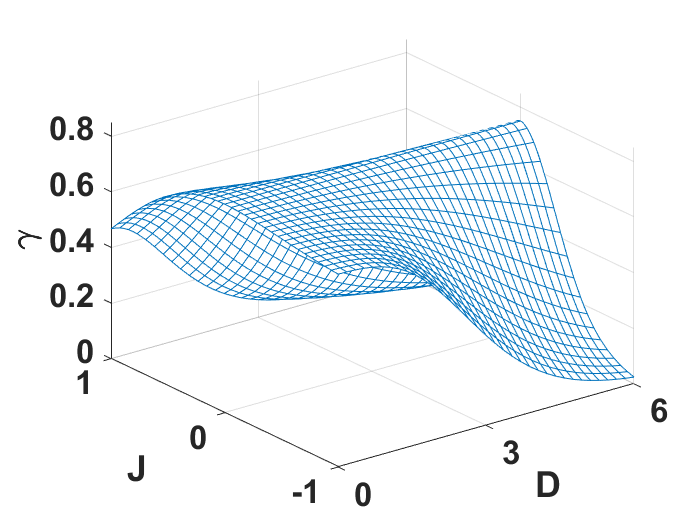}}}
    \caption*{Fig.2 Evolution of $C$ and $\gamma$ with $J$ and $D$}\label{Fig1}
\end{figure}

Fig.2 illustrates the dynamic evolution of $C$ and $\gamma$ with the real coupling coefficient $J$ and the value of the coupling strength of DM interaction $D$. One can see that the evolution trend is consistent with the Fig.1 in Ref.\cite{01}. For further investigation, we introduce $L$ as the left-hand side of uncertainty relation (10) and $L=\sum_{k=1}^K\mathrm {E}\left[\mathrm {V}\left(Q_k^A|O_k^{C}\right)\right]$. The dynamic evolution of $W$ and $L$ are shown in Fig.3.\\

\begin{figure}[H]
    \centering
    \setcounter{subfigure}{0}
    \subfigure[Evolution of $W$ and $L$ with $T$ when $J=1$, $D=1$ and $U_{tra}$ is taken as the upper bound of Eq.(10)]{%
        \resizebox{7cm}{5cm}{\includegraphics{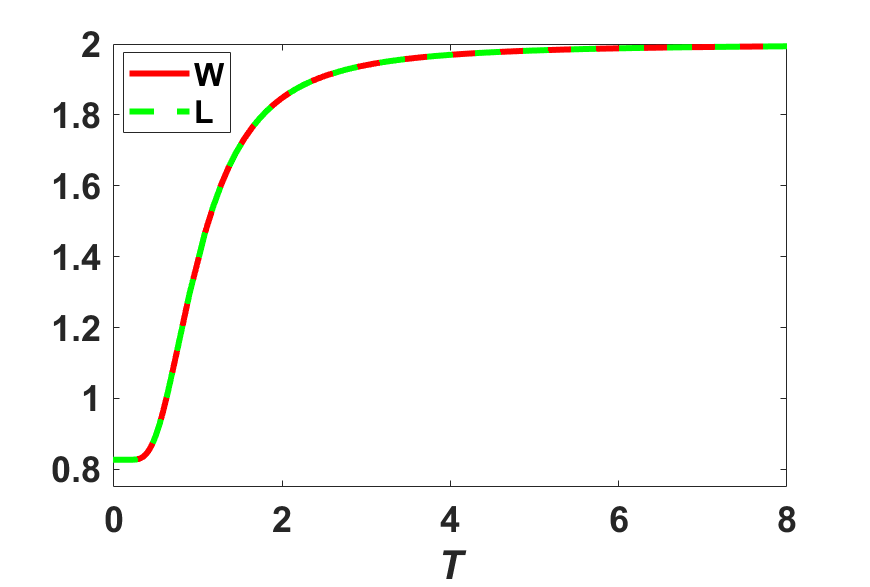}}}
    \hspace{5pt}
    \subfigure[Evolution of $W$ and $L$ with $T$ when $J=1$, $D=1$ and $U_{tra}$ is taken as the upper bound of Eq.(8) and (9)]{%
        \resizebox{7cm}{5cm}{\includegraphics{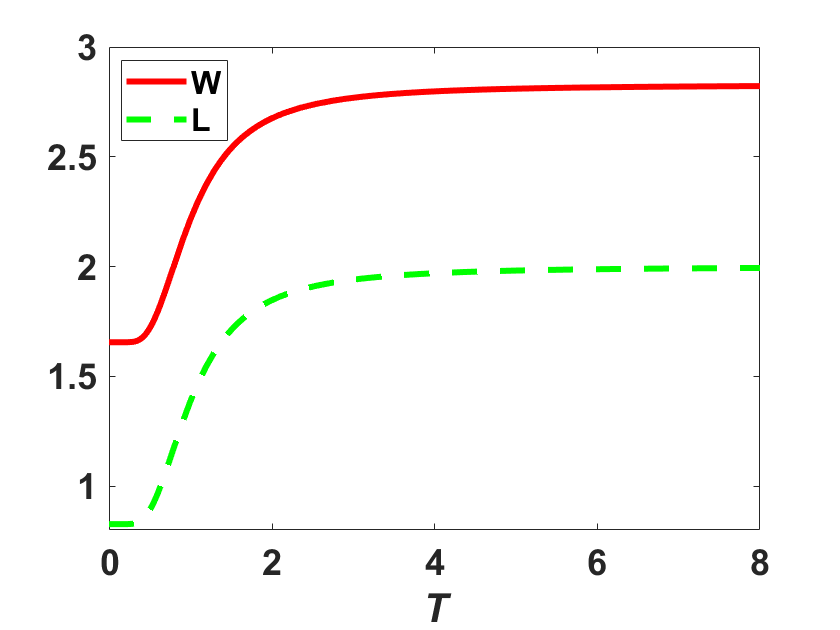}}}
    \caption*{Fig.3 Evolution of $W$ , $L$ with $T$ when $J=1$ and $D=1$}\label{Fig4}
\end{figure}

Fig.3 indicates that as the temperature increases, the left-hand side $L$ and upper bound $W$ also increase. In other words, the higher the system temperature is, the higher the uncertainty of the measurements is. Additionally, one can observe that $W$ is always precisely equal to $L$, indicating the tightness $U$ described by Eq.(23) is always a constant value 1  when $U_{tra}$ is taken as the upper bound of Eq.(10). The relevant further investigation is conducted in Fig.7, 8 and 9.\\

\begin{figure}[H]
    \centering
    \setcounter{subfigure}{0}
    \subfigure[Evolution of $W$ with $J$ and $D$ when $T=0.5$ and $U_{tra}$ is taken as the upper bound of Eq.(10)]{%
        \resizebox{6cm}{4cm}{\includegraphics{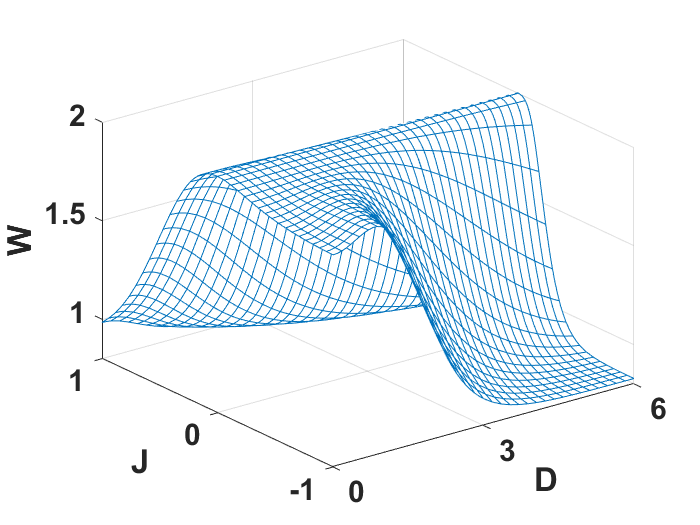}}}
    \hspace{5pt}
    \subfigure[Evolution of $W$ with $J$ and $D$ when $T=1$ and $U_{tra}$ is taken as the upper bound of Eq.(10)]{%
        \resizebox{6cm}{4cm}{\includegraphics{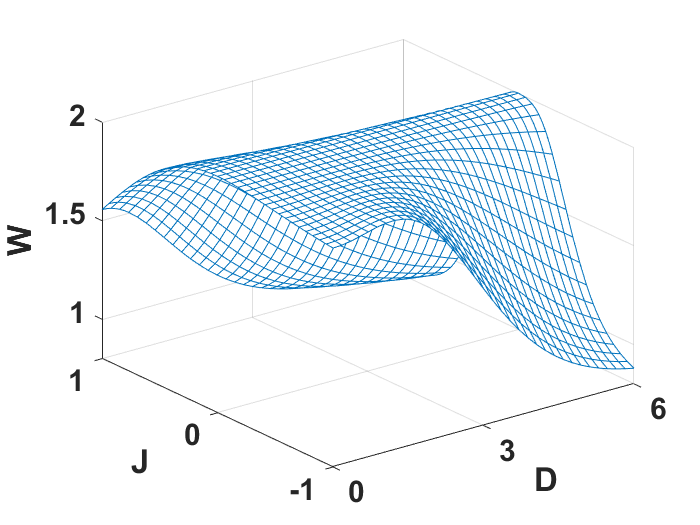}}}
    \\
    \subfigure[Evolution of $W$ with $J$ and $D$ when $T=0.5$ and $U_{tra}$ is taken as the upper bound of Eq.(8) and (9)]{%
        \resizebox{6cm}{4cm}{\includegraphics{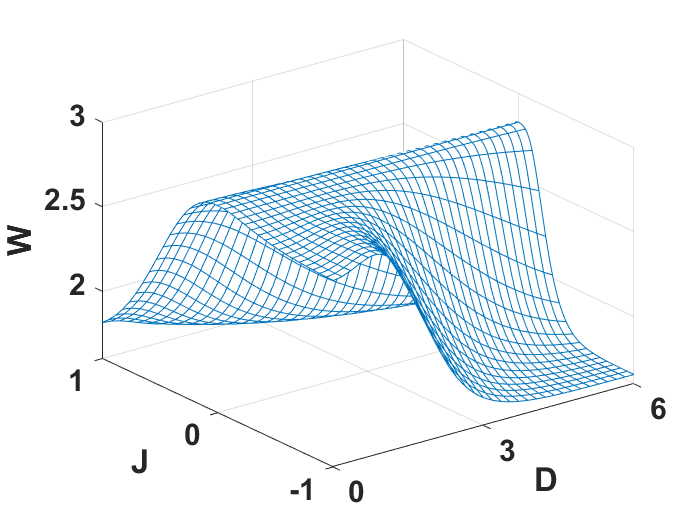}}}
    \hspace{5pt}
    \subfigure[Evolution of $W$ with $J$ and $D$ when $T=1$ and $U_{tra}$ is taken as the upper bound of Eq.(8) and (9)]{%
        \resizebox{6cm}{4cm}{\includegraphics{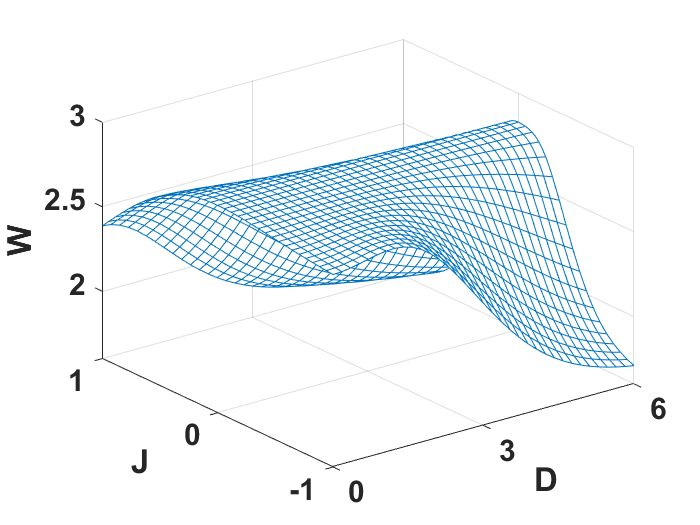}}}
    \caption*{Fig.4 Evolution of $W$ with $J$ and $D$ within different $T$ and $U_{tra}$}\label{Fig5}
\end{figure}

As shown in Fig.4, the dynamic evolution of the upper bound $W$ with $J$ and $D$ share similar evolution trends with the Fig.1 in Ref.\cite{01}. Based on all the figures above, one can see that the greater the entanglement is, the smaller the uncertainty of the measurement is. At the same time, the stronger mixedness indicates the stronger uncertainty. That is to say, $W$ and $C$ is roughly negatively correlated while $W$ and $\gamma$ is positively correlated. One can obtain that the mixedness is more closely related to uncertainty relation compared with the entanglement. Moreover, those figures indicate that the upper bound $W$ can be written as a function of $\gamma$ since $W$ and $\gamma$ are both related to $J$ and $D$. Therefore, we will perform further investigation about this relationship.\\

\begin{figure}[H]
    \centering
    \setcounter{subfigure}{0}
    \subfigure[Evolution of $W$, $C$ and $\gamma $ with $T$ when $J=1$, $D=1$ and $U_{tra}$ is taken as the upper bound of Eq.(10)]{%
        \resizebox{7cm}{5cm}{\includegraphics{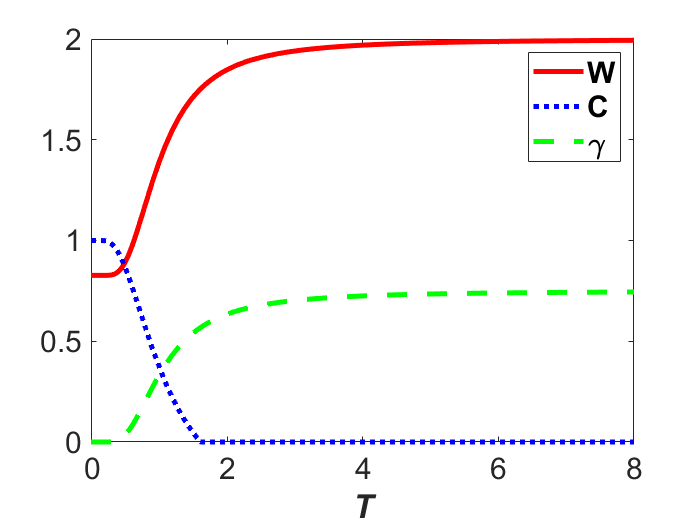}}}
    \hspace{5pt}
    \subfigure[Evolution of $W$, $C$ and $\gamma $ with $T$ when $J=1$, $D=1$ and $U_{tra}$ is taken as the upper bound of Eq.(8) and (9)]{%
        \resizebox{7cm}{5cm}{\includegraphics{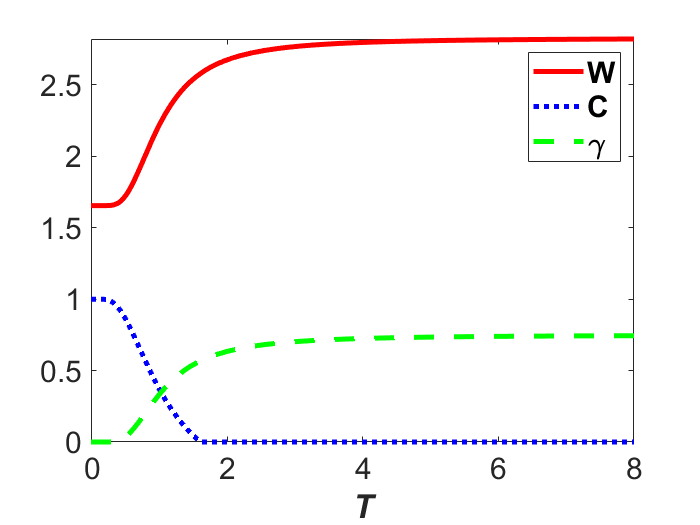}}}
    \caption*{Fig.5 Evolution of $W$, $C$ and $\gamma $ with $T$ when $J=1$, $D=1$}\label{Fig6}
\end{figure}

Fig.5 demonstrates that $W$ and $\gamma$ increase when the temperature become higher. But the higher the temperature is, the slower the increase of $\gamma$ and $W$ shows. In contrast, temperature $T$ and concurrence $C$ is negatively correlated and $C$ approaches to 0 when the temperature is high enough, which means the two-body entanglement disappears in high temperature situation. From Fig.5, one can see that compared with the entanglement, the relationship between the mixedness and uncertainty relation is closer, showing a more obvious single-value relationship. Then, we performed the further investigation of this single-value relationship between $W$ and $\gamma$ in Fig.6.\\

\begin{figure}[H]
    \centering
    \setcounter{subfigure}{0}
    \subfigure[Evolution of $W$ with $D$ and $\gamma$ when $J=1$ and $U_{tra}$ is taken as the upper bound of Eq.(10)]{%
        \resizebox{6cm}{4cm}{\includegraphics{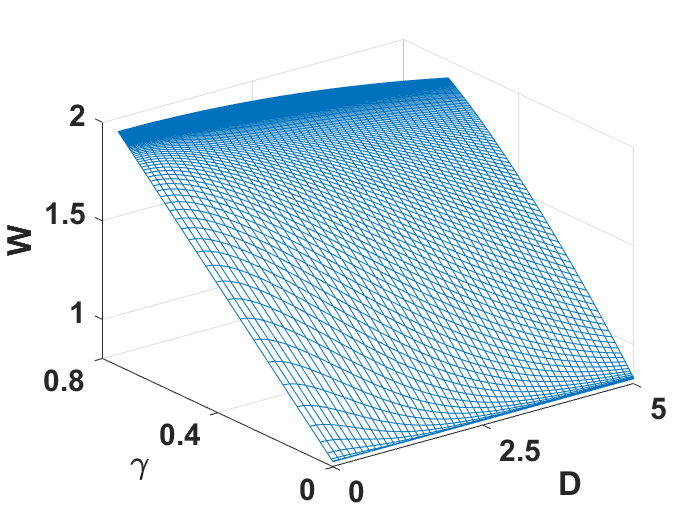}}}
    \hspace{5pt}
    \subfigure[Evolution of $W$ with $D$ and $\gamma$ when $J=1$ and $U_{tra}$ is taken as the upper bound of Eq.(8) and (9)]{%
        \resizebox{6cm}{4cm}{\includegraphics{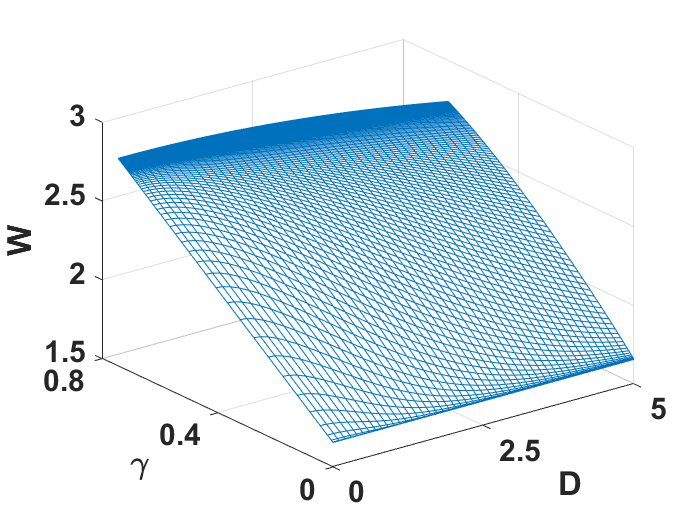}}}
    \\
    \subfigure[Evolution of $W$ with $D$ and $\gamma$ when $J=-1$ and $U_{tra}$ is taken as the upper bound of Eq.(10)]{%
        \resizebox{6cm}{4cm}{\includegraphics{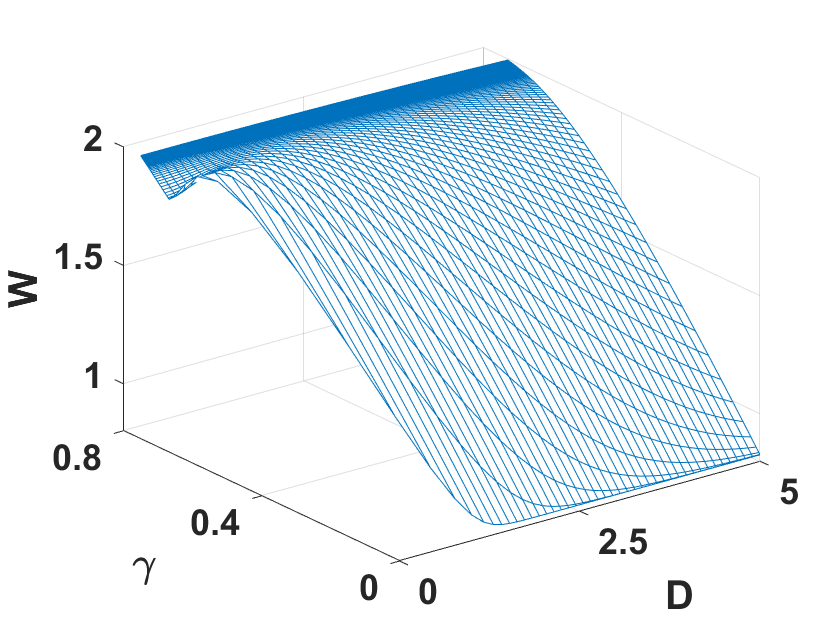}}}
    \hspace{5pt}
    \subfigure[Evolution of $W$ with $D$ and $\gamma$ when $J=-1$ and $U_{tra}$ is taken as the upper bound of Eq.(8) and (9)]{%
        \resizebox{6cm}{4cm}{\includegraphics{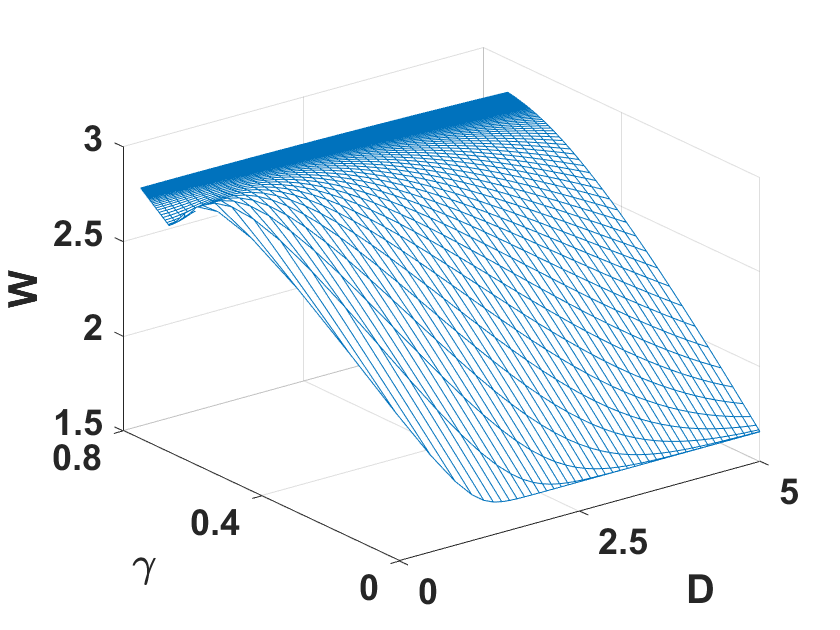}}}
    \caption*{Fig.6 Evolution of $W$ with $D$ and $\gamma$ within different $J$ and $U_{tra}$}\label{Fig7}
\end{figure}

Fig.6 illustrates that there do exist an obvious single-value relationship between the mixedness $\gamma$ and the upper bound $W$ of the reverse uncertainty relation.
It should be noted that the corresponding functional form is determined only by the sign of the real coupling coefficient $J$, and is independent of the specific value of $J$, as illustrated in Fig.6.
Also, from the single-value relationship between $W$ and $\gamma$, one can see that the upper bound of the uncertainty relation tends to its minimum as the mixedness of the system approaches the minimum value. That is to say, the reverse uncertainty relation reaches its optimal performance when the system is in pure state. Furthermore, as $\gamma$ approaches 0, the uncertainty relation is observed to be better, signifying that a reduction in the system's mixedness correlates with increasing uncertainty relation. As $\gamma$ ascends, the system's uncertainty relation become worse, indicating that increased mixedness is directly associated with worse uncertainty relation. Ultimately, a single-value relationship is obvious between the system's mixedness $\gamma$ and its upper bound $W$ of the uncertainty relation (11), which demonstrates a high degree of correlation. Additionally, the single-value relationship between the mixedness and the uncertainty relation of the normal uncertainty relation in the Heisenberg model with DM interaction for the incompatible measurements has also been observed in the Fig.3 in Ref.\cite{01}.\\



\begin{figure}[H]
    \centering
    \setcounter{subfigure}{0}
    \subfigure[Evolution of $U$ with $D$ and $J$ when $U_{tra}$ is taken as the upper bound of Eq.(10)]{%
        \resizebox{6cm}{4cm}{\includegraphics{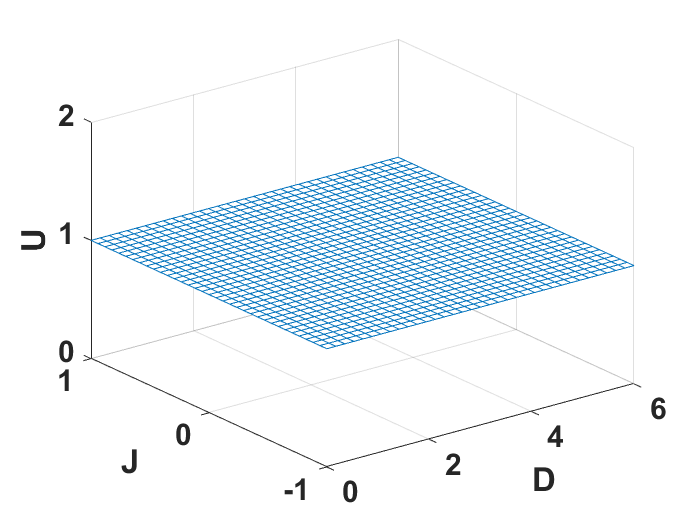}}}
    \hspace{5pt}
    \subfigure[Evolution of $U$ with $D$ and $J$ when $U_{tra}$ is taken as the upper bound of Eq.(8) and (9)]{%
        \resizebox{6cm}{4cm}{\includegraphics{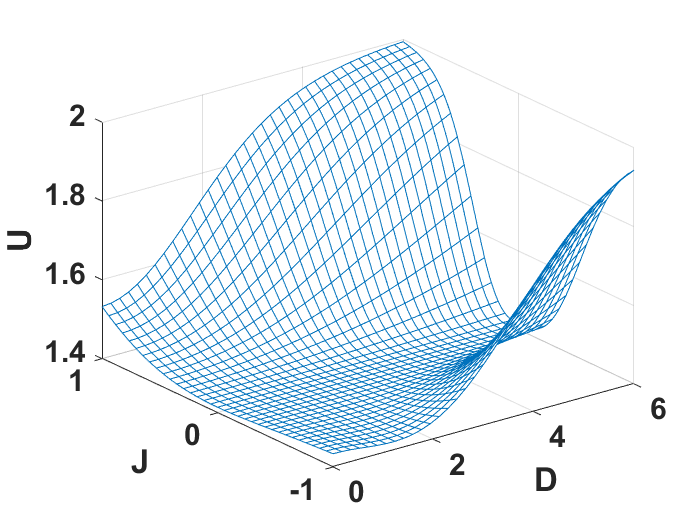}}}
    \caption*{Fig.7 Evolution of $U$ with $D$ and $J$}\label{Fig8}
\end{figure}

Fig.7 (a) shows that $U$ always approaches its optimal performance when $U_{tra}$ is taken as the upper bound of Eq.(10). $U$ is always a constant value 1 in this case, suggesting the tightness is independent of $J$, $D$ and always highly optimal. But when $U_{tra}$ is taken as the upper bound of Eq.(8) and (9), one can see that $U$ is roughly positive correlated to $J$ and $D$ according to Fig.7 (b). It indicates that $U$ has relationship with $\gamma$ as $\gamma$ is also related to $J$ and $D$. On one hand, as $D$ descends, $U$ descends too. In other words, the tightness of the system will be tighter when the coupling strength of DM interaction weakens. On the other hand, $U$ is observed to approaches to 1 when $J$ approaches to 0. That is to say, as the coupling strength weakens, the tightness will tend to be optimal. The dynamic evolution of the tightness illustrated in Fig.7 is similar to the Fig.4 in Ref.\cite{01}.\\

\begin{figure}[H]
    \centering
    \setcounter{subfigure}{0}
    \subfigure[Evolution of $U$, $C$ and $\gamma$ with $T$ when $J=1$, $D=1$ and $U_{tra}$ is taken as the upper bound of Eq.(10)]{%
        \resizebox{7cm}{5cm}{\includegraphics{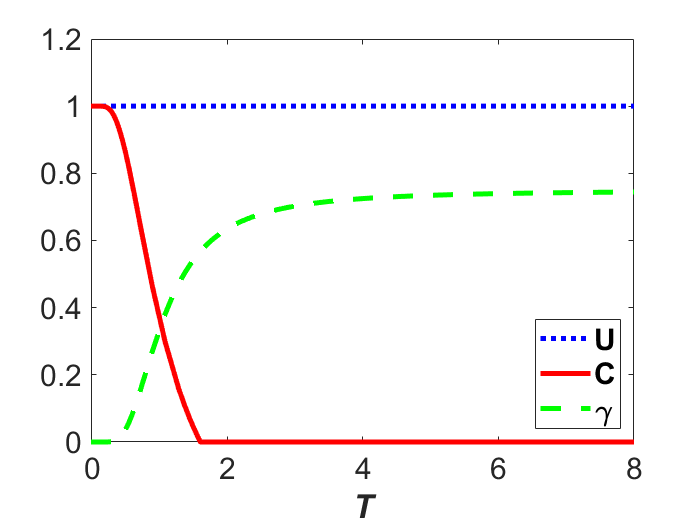}}}
    \hspace{5pt}
    \subfigure[Evolution of $U$, $C$ and $\gamma$ with $T$ when $J=1$, $D=1$ and $U_{tra}$ is taken as the upper bound of Eq.(8) and (9)]{%
        \resizebox{7cm}{5cm}{\includegraphics{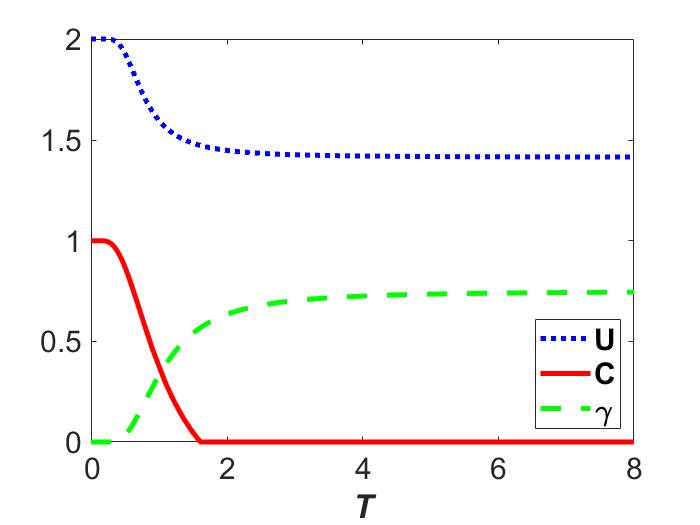}}}
    \caption*{Fig.8 Evolution of $U$ , $C$ and $\gamma$ with $T$ when $J=1$ and $D=1$}\label{Fig9}
\end{figure}

It can be obtained that $U$ is always optimal when $U_{tra}$ is taken as the upper bound of Eq.(10) from Fig.8 (a), indicating that the mixedness and the tightness of the system are independent in this situation. From Fig.8 (b), a negative correlation between $U$ and $\gamma$ can be observed, and a positive correlation between $U$ and $C$ can also be obtained when $U_{tra}$ is taken as the upper bound of Eq.(8) and (9). The relationship between $U$ and $\gamma$ shows a higher degree of correlation than the one between $U$ and $C$. That is to say, Fig.8 suggests that there exists a single-value relationship between the tightness and the mixedness of the system as same as the one between the uncertainty relation and the mixedness of the system. We can also obtain that the trend of the new relationship between $U$ and $\gamma $ should be roughly opposite.\\

\begin{figure}[H]
    \centering
    \setcounter{subfigure}{0}
    \subfigure[Evolution of $U$ with $D$ and $\gamma$ when $J=1$ and $U_{tra}$ is taken as the upper bound of Eq.(10)]{%
        \resizebox{6cm}{4cm}{\includegraphics{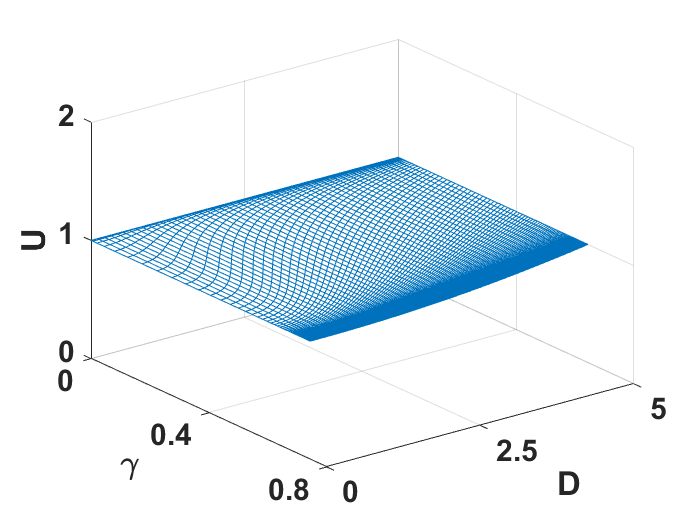}}}
    \hspace{5pt}
    \subfigure[Evolution of $U$ with $D$ and $\gamma$ when $J=1$ and $U_{tra}$ is taken as the upper bound of Eq.(8) and (9)]{%
        \resizebox{6cm}{4cm}{\includegraphics{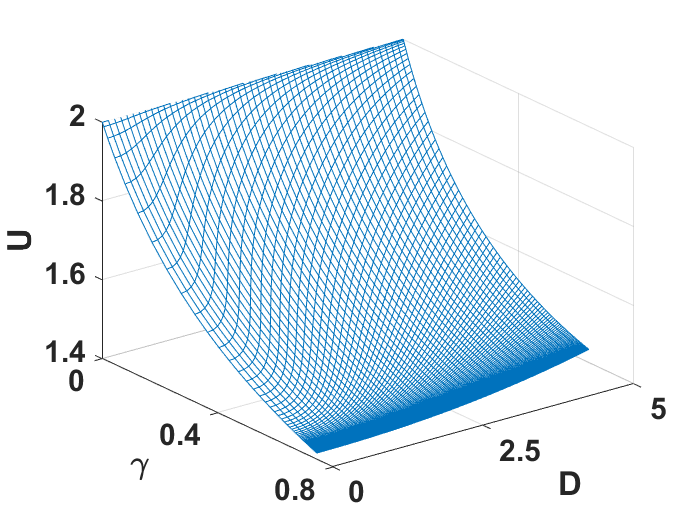}}}
    \\
    \subfigure[Evolution of $U$ with $D$ and $\gamma$ when $J=-1$ and $U_{tra}$ is taken as the upper bound of Eq.(10)]{%
        \resizebox{6cm}{4cm}{\includegraphics{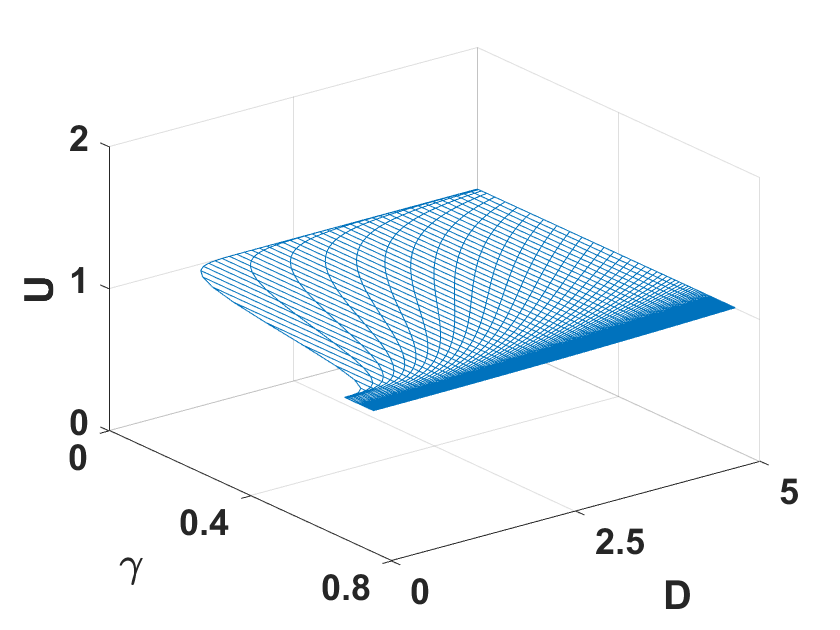}}}
    \hspace{5pt}
    \subfigure[Evolution of $U$ with $D$ and $\gamma$ when $J=-1$ and $U_{tra}$ is taken as the upper bound of Eq.(8) and (9)]{%
        \resizebox{6cm}{4cm}{\includegraphics{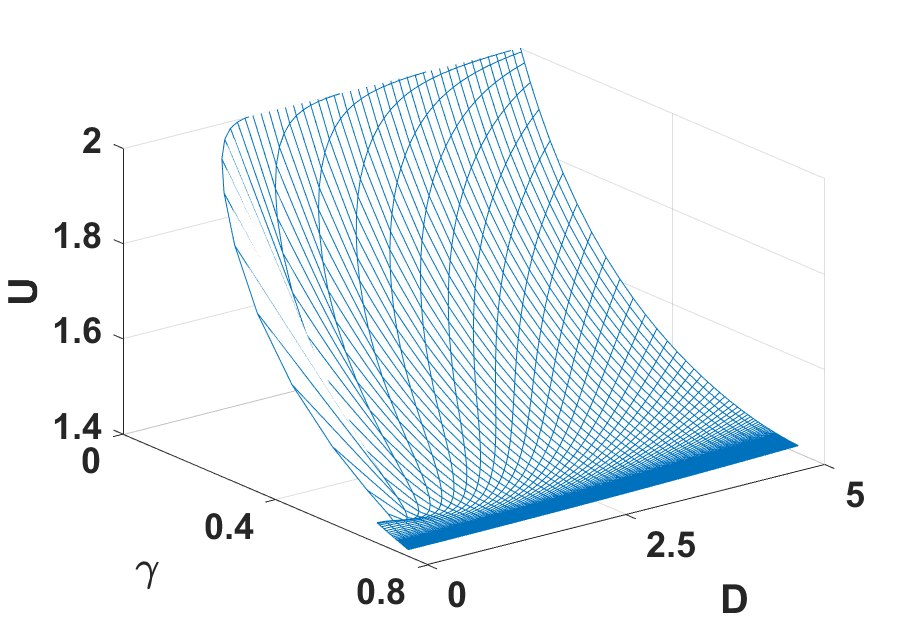}}}
    \caption*{Fig.9 Evolution of $U$ with $D$ and $\gamma$ within different $J$ and $U_{tra}$}\label{Fig10}
\end{figure}

Fig.9 (a) and (c) demonstrates that $U$ is always a constant value when $U_{tra}$ is taken as the upper bound of Eq.(10). That is to say, the tightness of the system has no relationship with the parameters, namely $\gamma$, $D$ and $T$ of the system and always reachs its optimal state in this situation. Fig.9 (b) and (d) shows that $U$ increases as $\gamma$ approaches 0 when $U_{tra}$ is taken as the upper bound of Eq.(8) and (9). That is to say, the tightness of the system approaches to its optimal performance as the mixedness of the system increases. On the contrary, $U$ descends while $\gamma$ ascends, showing that the greater the mixedness of the system is, the tighter the tightness of the system is. In conclusion, there exists a single-value relationship between the mixedness $\gamma$ and the tightness $U$, which shows a high degree of correlation with the Fig.6 in Ref.\cite{01}.\\

\section{conclusion} \label{C4}
In conclusion, we here construct a new quantum-control-assisted reverse uncertainty relation, which indicates that the upper bound of the reverse uncertainty relation without conditional system can be broken with the help of quantum control system. The  investigation reveals that the uncertainty of the incompatible measurements is negatively correlated to the mixedness of the system and positively correlated to the entanglement. It is found that the relationship between mixedness and uncertainty is closer than that between entanglement and uncertainty, and there exists an interesting single-value relationship between the mixedness and the tightness and upper bound of the uncertainty relation. \textbf{Concisely, this uncertainty relation demonstrates that the greater the mixedness is, the greater the uncertainty of the measurement results and the tighter the tightness of the system is. This single-value relationship suggests that the tightness and upper bound of the uncertainty relation can be expressed as the functional form of the mixedness of the system, offering a possible direction for further research.} Also, we show that the single-value relationship is a common nature of the reverse uncertainty relation and the normal one.\\

\section*{Acknowledgments}
This work was supported by the National Natural Science Foundation of China (Grant Nos. 12474353, 12074027)  and the Fundamental Research Funds for the Central Universities. 

\section*{Author contributions}

 Xiao Zheng and Guo-Feng Zhang conceived and designed the research. Qiyi Li and Shao-Qiang Ma performed analysis and wrote the manuscript. Qiyi Li prepared all the figures. Sansheng Wang commented on the manuscript. All authors discussed the results and revised the manuscript.

\end{document}